\documentclass[pra,twocolumn,showpacs,superscriptaddress]{revtex4-1}
\usepackage{amsmath,amscd,amssymb,color}
\usepackage{graphicx,amsfonts,epsf}
\usepackage{epstopdf}
\usepackage{float}
\usepackage{hyperref}
\usepackage{enumerate}
\newcommand{\ie}{\textit{i.e.~}}
\usepackage{ulem}

\begin{document}
\title{Efficient experimental characterization of quantum processes 
via compressed sensing on an NMR quantum processor}
\author{Akshay Gaikwad}
\email{ph16010@iisermohali.ac.in}
\affiliation{Department of Physical Sciences, Indian
Institute of Science Education \& 
Research Mohali, Sector 81 SAS Nagar, 
Manauli PO 140306 Punjab India.}
\author{Arvind}
\email{arvind@iisermohali.ac.in}
\affiliation{Department of Physical Sciences, Indian
Institute of Science Education \& 
Research Mohali, Sector 81 SAS Nagar, 
Manauli PO 140306 Punjab India.}
\affiliation{Vice Chancellor, Punjabi University Patiala,
Punjab 147002, India.}
\author{Kavita Dorai}
\email{kavita@iisermohali.ac.in}
\affiliation{Department of Physical Sciences, Indian
Institute of Science Education \& 
Research Mohali, Sector 81 SAS Nagar, 
Manauli PO 140306 Punjab India.}
\begin{abstract}
We employ the compressed sensing (CS) algorithm and a
heavily reduced data set to experimentally perform true
quantum process tomography (QPT) on an NMR quantum
processor.  We obtain the estimate of the process matrix
$\chi$ corresponding to various two- and three-qubit quantum
gates with a high fidelity.  The CS algorithm is implemented
using two different operator bases, namely, the standard
Pauli basis and the Pauli-error basis.  We experimentally
demonstrate that the performance of the CS algorithm is
significantly better in the Pauli-error basis, where the
constructed $\chi$ matrix is maximally sparse. We compare
the standard least square (LS) optimization QPT method with
the CS-QPT method and observe that, provided an appropriate
basis is chosen, the CS-QPT method performs  significantly
better as compared to the LS-QPT method.  In all the cases
considered, we obtained experimental fidelities greater than
0.9 from a reduced data set, which was approximately five to
six times smaller in size than a full data set.  We also
experimentally characterized the reduced dynamics of a
two-qubit subsystem embedded in a three-qubit system, and
used the CS-QPT method to characterize processes
corresponding to the evolution of two-qubit states under
various $J$-coupling interactions.
\end{abstract} 
\maketitle 
\section{Introduction}
\label{intro}
An essential task in experimental quantum information
processing is the characterization of quantum states and
their dynamics, which is typically achieved via quantum
state tomography (QST)~\cite{li-pra-2017} and quantum
process tomography (QPT)\cite{chuang-jmo-09}.  The 
experimental resources
required  to implement QST and QPT 
grow exponentially with the size of
system, which makes these methods infeasible beyond a few
qubits\cite{mohseni-pra-2008}.  Hence developing techniques
to reduce the experimental resources required for quantum
process tomography is of paramount importance in scaling up
quantum technologies.  Several strategies have been designed
to address these issues, such as methods based on the
least-square (LS) linear inversion
technique\cite{miranowicz-pra-2014}, linear regression
estimation\cite{qi-quantum-inf-2017}, maximum likelihood
estimation (MLE)\cite{james-pra-2001},  self-guided
tomography\cite{rambach-prl-2021} and numerical
strategies\cite{kaznady-pra-2009}.  Several QST protocols
have been extended to perform QPT, which include MLE-based
QPT\cite{obrien-prl-04}, LS-based
QPT\cite{trystan-arxiv-2021}, simplified
QPT\cite{kosut-njp-2009}, convex optimization-based
QPT\cite{jin-2019}, selective and efficient QPT
\cite{perito-pra-2018}, adaptive
QPT\cite{pogorelov-pra-2017}, and ancilla-assisted
QPT\cite{altepeter-prl-2003}.  These protocols have been
successfully demonstrated on various physical systems such
as
NMR~\cite{gaikwad-pra-2018,xin-npj-2019,xin-phys-app-2020,Gaikwad-qip-2021,zhao-pra-2021},
NV-centers\cite{zhang-prl-2014}, linear
optics\cite{paz-prl-2010}, superconducting
qubits\cite{neeley-nature-2008,
chow-prl-2009,gaikwad-ijqi-2020} and ion trap-based quantum
processors\cite{riebe-prl-2006}.

Methods such as Monte-Carlo process
certification\cite{silva-prl-2011} and randomized
benchmarking\cite{knill-pra-2008} have been developed to
address scalability issues in standard QST and QPT methods.
However, they are limited in scope as they do not provide
the full process matrix and hence cannot be used to identify
gate errors or improve gate fidelity.  Other methods such as
ancilla-assisted QPT are able to significantly reduce the
experimental complexity, however the issues of scalability
remain. The CS algorithm borrows ideas from classical
signal processing which posits that even heavily
undersampled sparse signals can be efficiently
reconstructed. The CS algorithm relies on reformulating QST
and QPT tasks as a constrained convex optimization problem,
and is able to perform complete and true characterization of
a given quantum process from a heavily reduced data set
without performing actual projective measurements, and does
not need any extra resources such as ancilla qubits.  CS-QST
and CS-QPT have been successfully used to reconstruct
unknown quantum states from NMR data~\cite{yang-pra-2017},
to characterize quantum gates based on superconducting Xmon
and phase qubits~\cite{rodionov-prb-2014}, and to perform
efficient estimation of process matrices of  photonic
two-qubit gates~\cite{shabani-prl-2011}.

In this work we utilize the CS algorithm to perform QPT of
various two- and three-qubit quantum gates on an NMR quantum
processor.  We also demonstrate the efficacy of the CS-QPT
protocol in characterizing two-qubit dynamics in a
three-qubit system.  We experimentally estimate the full
process matrix corresponding to a given quantum process with
a high fidelity, from a drastically reduced set of initial
states and outcomes.  The CS-QPT algorithm is able to
efficiently characterize a given quantum process provided
the corresponding process matrix is sufficiently sparse (\ie
most of its matrix elements are zero).  We use two different
operator basis sets to estimate the process matrix using the
CS algorithm, namely,  the standard Pauli basis and the
Pauli-error basis (where the process matrix is maximally
sparse~\cite{rodionov-prb-2014}, \ie it contains only one
non-zero element).  We also compare the performance of the
CS-QPT and the LS-QPT methods using significantly reduced
data sets in both the standard Pauli basis and the
Pauli-error basis.  We obtained experimental fidelities of
greater than 0.9 from a reduced data set of a size
approximately 5 to 6 times smaller than the size of a full
data set, and our results indicate that the CS-QPT method is
significantly more efficient than standard QPT methods.

This paper is organized as follows: In Section~\ref{qpt} we
detail the implementation of the CS algorithm in the context
of QPT.  The standard QPT protocol is briefly described
in Section~\ref{sec2.1}, while the CS-QPT method is
given in Section~\ref{sec2.2}.
Section~\ref{nmr-expt} describes the experimental
implementation of the CS-QPT methods using two and three
NMR qubits.
In Sections~\ref{csqpt2q} and \ref{csqpt3q},
we present the quantum
circuit and the corresponding NMR implementation of the
CS-QPT method for two- and three-qubit quantum gates, respectively. 
Section~\ref{sec3.3} contains a description of the CS-QPT
implementation to capture
two-qubit quantum dynamics embedded in a three-qubit system.
Section~\ref{sec3.4} contains a comparison  of the
CS-QPT and LS-QPT protocols.
Section~\ref{concl} contains a few concluding remarks.  
\section{QPT for a reduced data set}  
\label{qpt}
\subsection{Standard QPT and $\chi$ matrix representation}
\label{sec2.1}
In a fixed basis set $\lbrace E_i \rbrace$, a quantum map
(a completely positive map)
$\Lambda$ can be written as~\cite{kraus-book-1983}:
\begin{equation}
\Lambda(\rho) = \sum_{m,n} \chi_{mn} E_m \rho E_n^{\dagger}
\label{eq2}
\end{equation}
where the Kraus operators are expanded as $A_i =
\sum_{k}a_{ik}E_k$ and the quantities
$\chi_{mn}=\sum_{i}a_{im}a_{in}^*$ are the elements of the
process matrix $\chi$ characterizing the quantum map
$\Lambda$.  In a $d$-dimensional Hilbert space, $\chi$ is a
$d^2 \times d^2$ dimensional positive semi-definite matrix
and $d^4$ real independent parameters are required to
uniquely represent it. The number of required parameters
reduces from $d^4$ to ($d^4-d^2$) for trace preserving
processes~\cite{obrien-prl-04}.

The standard QPT protocol estimates the complete $\chi$
matrix by preparing the system in different quantum states,
letting it evolve under the given quantum process, and then
measuring a set of observables~\cite{childs-pra-2001}. The
full data set for QPT can be acquired using tomographically
complete sets of input states $\lbrace \rho_1,
\rho_2,....,\rho_k \rbrace $, letting them
undergo the desired quantum process $\chi$, and 
measuring an observable $M$ from the set of measurement operators
$\lbrace M_1, M_2,..., M_l \rbrace$, leading to:
\begin{equation}
B^i_j=\text{Tr}(M_j\Lambda(\rho_i)) = 
\sum_{m,n} \chi_{mn} \text{Tr}(M_j E_m \rho_i E_n^{\dagger})
\label{eq3}
\end{equation}
For all input states $\lbrace \rho_i \rbrace$ 
and measurement operators $\lbrace M_j \rbrace$ 
in Eq.~(\ref{eq3}), the relationship between
the vector of outcomes and the true process matrix
 can be rewritten in 
a compact form~\cite{childs-pra-2001}:
\begin{equation}
\overrightarrow{B}(\chi) = \Phi \overrightarrow{\chi}
\label{eq4}
\end{equation} 
where $\overrightarrow{B}(\chi)$ and $\overrightarrow{\chi}$
are vectorized forms of $B^i_j$ and $\chi_{mn}$
respectively, and $\Phi$ is the coefficient matrix with the
entries $\Phi_{ji,mn} = \text{Tr}(M_j E_m \rho_i
E_n^{\dagger})$. 

We note here that using the standard QPT method  may not
always lead to a positive semi-definite experimentally
constructed $\chi$ matrix, due to experimental
uncertainties. This problem can be resolved by reformulating
the linear inversion problem as a constrained convex
optimization problem as follows\cite{Gaikwad-qip-2021}:
 
\begin{subequations} 
\begin{alignat}{2}
\min_{\chi}\quad & \Vert
\overrightarrow{B}^{exp}-\overrightarrow{B}(\chi)\Vert_{l_2}\label{eq5}\\
\text{subject to}\quad       & \chi \geq 0,\label{eq5:constraint1}\\
        &
\sum_{m,n}\chi_{mn}E_m^{\dagger}E_n =
I_d.\label{eq5:constraint2} \end{alignat} 
\end{subequations}
where the vector $\overrightarrow{B}^{exp}$ is constructed
using experimental measurement outcomes.  This method is
referred to as the least square (LS) optimization method. In
this work, we study the performance of the LS-QPT method for
a reduced data set.
\subsection{Compressed sensing QPT} 
\label{sec2.2}
Compressed sensing methods work well if the process matrix
is sparse in some known basis and
rely on compressing the information 
contained in a process of large size into 
one of much smaller size and use 
efficient convex optimization algorithms to 
``unpack'' this compressed information.
The
CS-QPT method hence provides a way to reconstruct the complete and
true $\chi$ matrix of a given quantum process from a drastically
reduced data set, provided that the $\chi$ matrix is
sufficiently sparse in some known basis \ie, the number of
non-zero entries in the $\chi$ matrix is small.
It is to be noted that the sparsity is a property of the
map representation and not the map itself.
Specifically, for quantum gates which are trace-preserving
unitary quantum processes, one can always find the proper
basis in which the corresponding $\chi$ matrix is maximally
sparse~\cite{korotkov-arxiv-2013,kosut-arxiv-2008}.

Estimating a sparse process matrix with an unknown
sparsity pattern from an underdetermined set of
linear equations can be done using numerical optimization
techniques.
For trace-preserving maps, the complete
convex optimization problem for CS-QPT is formulated as
follows: 
\begin{subequations} 
\begin{alignat}{2}
\!\min_{\chi}        \quad&
\Vert{\overrightarrow{\chi}}\Vert_{l_1}\label{eq6}\\
\text{subject to}\quad       & \Vert
\overrightarrow{B}^{exp}-\Phi \overrightarrow{\chi}
\Vert_{l_2} \leq \epsilon,\label{eq6:constraint1}\\ &
\chi \geq 0,\label{eq6:constraint2}\\ &
 \sum_{m,n}\chi_{mn}E_m^{\dagger}E_n =
I_d.\label{eq6:constraint3} 
\end{alignat} 
\end{subequations}
where Eq.~(\ref{eq6}) is the main objective function which is
to be minimized and Eq.~(\ref{eq6:constraint1}) is the
standard constraint involved in the CS algorithm;
Eq.~(\ref{eq6:constraint2}) and Eq.~(\ref{eq6:constraint3}) denote
the positivity and trace preserving constraints of the
process matrix, respectively. The parameter $\epsilon$
quantifies the level of uncertainty in the measurement, \ie
the quantity $\overrightarrow{B}^{{\rm exp}}=\Phi
\overrightarrow{\chi}_0+\overrightarrow{z}$ is observed,
with $\Vert \overrightarrow{z} \Vert_{l_2}  \leq \epsilon$,
where $\overrightarrow{\chi}_0$ is the vectorized form of
the true process matrix and $\overrightarrow{z}$ is an unknown
noise vector. The general $l_p$- norm of
a given vector $ {\overrightarrow{x}}$ is defined as:
$\|x\|_{p}=\left(\sum_{i}\left|x_{i}\right|^{p}\right)^{1 /
p}$. If the process matrix is sufficiently sparse and the
coefficient matrix $\Phi$ satisfies the restricted isometry
property (RIP) condition, then by solving the optimization
problem delineated in Eq.~(\ref{eq6}), one can accurately
estimate the process matrix~\cite{shabani-prl-2011}.
The RIP condition is satisfied if the coefficient
matrix $\Phi$ satisfies the following
conditions~\cite{shabani-prl-2011,rodionov-prb-2014}:
\begin{itemize} \item[(i)] \begin{equation} 1-\delta_s \leq
\frac{\Vert \Phi \overrightarrow{\chi}_1 - \Phi
\overrightarrow{\chi}_2 \Vert_{l_2}^2}
{\Vert\overrightarrow{\chi}_1 - \overrightarrow{\chi}_2
\Vert_{l_2}^2}\leq 1+\delta_s \label{rip1} \end{equation}
for all $s$-sparse vectors $\overrightarrow{\chi}_1$ and
$\overrightarrow{\chi}_2$.  
An $N \times 1$ dimensional
vector $\overrightarrow{x}$ is $s$-sparse, if only $s < N$
elements are non-zero.  
\item [(ii)] The value of the
isometry constant $\delta_s < \sqrt{2}-1$.
The restricted isometry constant (RIC) of a
matrix $A$ measures how close to an isometry is the action of
$A$ on vectors with a few nonzero entries, measured in the
$l_2$-norm~\cite{emmanuel-crm-2008}. 
Specifically, the upper and lower RIC of a
matrix $A$ of size $n \times N$ is the maximum and the minimum
deviation from unity (one) of the largest and smallest,
respectively, square of singular values of $\text { all
}\left(\begin{array}{c} N \\ k \end{array}\right) \text {
matrices }$ formed by taking $k$ columns from $A$. 
\item[(iii)] The size of the data set is sufficiently large
\ie   $m_{\text{conf}}\geq C_0 s  \text{log}(d^4/s)$ where
$C_0$ is a constant,  
$m_{\text{conf}}$ is the size of the data set, $s$ is 
the sparsity of
the process matrix and $d$ is the dimension of 
the Hilbert space. 
\end{itemize} 

Once the basis operators $\lbrace E_{\alpha}
\rbrace$ and the configuration space  $\lbrace \rho_i,
M_j\rbrace $ are chosen, the coefficient matrix
$\Phi_{\text{full}}$ corresponding to the entire data set is
fully defined and does not depend on the measurement
outcomes. It has been shown that if $\Phi_{m}$ is built by
randomly selecting $m$ rows (\ie $m$ number of random
configurations) from $\Phi_{\text{full}}$ then it is most
likely to satisfy the RIP
conditions~\cite{rodionov-prb-2014}.  
Hence the sub-matrix
$\Phi_m \in \Phi_{\text{full}} $ together with the
corresponding observation vector $\overrightarrow{B}^{exp}_m
\in \overrightarrow{B}^{exp}_\text{full}$ can be used to
estimate the process matrix by solving the optimization
problem (Eq.~(\ref{eq6})). 
  
In this study, we use two different operator basis sets,
namely the standard Pauli basis (PB) and the Pauli-error
basis (PEB).  For both bases, the orthogonality condition is
given by $ \langle E_{\alpha} \vert E_{\beta} \rangle = d
\delta_{\alpha \beta} $.  
For
an $n$-qubit system, 
the basis operators $P_i$ in the PB set are 
$P_i = \lbrace I, \sigma_x, \sigma_y, \sigma_z \rbrace
^{\otimes n}$, while the basis operators $E_i$ in 
the PEB set are:
$E_i = UP_i$, where $U$ is the desired unitary matrix for
which the process matrix needs to be estimated.  Furthermore,
the process matrix in PEB corresponding to the desired $U$,
is always maximally sparse, \ie   it contains only one
non-zero element~\cite{rodionov-prb-2014}.  The convex
optimization problems involved in LS-QPT and CS-QPT
(Eq.~(\ref{eq5}) and Eq.~(\ref{eq6}), respectively) can be solved
efficiently using the YALMIP\cite{lofberg-2004} MATLAB
package, which employs SeDuMi\cite{sturm-oms-1999} as a
solver.
\begin{figure}
\includegraphics[angle=0,scale=1]{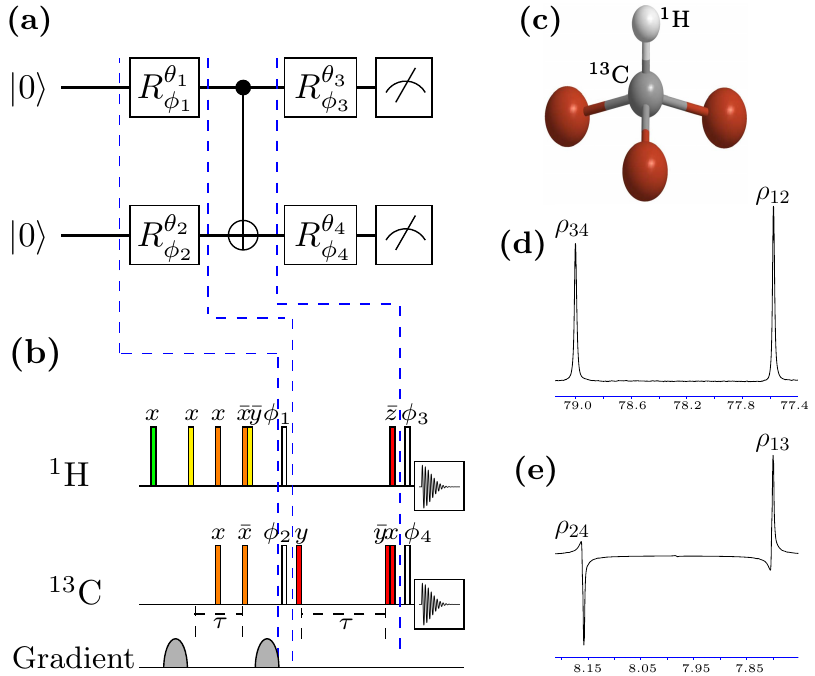} 
\caption{(Color online) (a) Quantum circuit
to implement CS-QPT of a CNOT gate.  Single-qubit unitary
operations $R_{\phi}^\theta$  are achieved via rotations by
an angle $\theta$ and phase $\phi$. The first 
block represents the preparation of the desired input state, while
the second and third blocks represent the quantum
process corresponding to the CNOT gate and the measurement,
respectively. (b) NMR implementation of the quantum circuit
given in panel (a). The rectangles filled with red, orange,
yellow and green color denote $\frac{\pi}{2}$, $\pi$,
$\frac{\pi}{4}$ and $\frac{\pi}{3}$ pulses, respectively, 
with the rf phase
written above each pulse. 
The unfilled rectangles with
the phases $\phi_1$, $\phi_2$, $\phi_3$, and $\phi_4$
represent pulses with flip angles $\theta_1$,
$\theta_2$, $\theta_3$, and $\theta_4$, respectively.
The gradient line denotes 
$z$-gradient pulses. The evolution time
period $\tau=\frac{1}{2J_{CH}}$ where $J_{CH}$ is the scalar
coupling constant. (c) ${}^{13}$C-labeled chloroform molecule
with ${}^{1}$H and ${}^{13}$C labeling the
first and second qubits, respectively. (d) and (e) depict
the
NMR spectra of ${}^{13}$C and ${}^{1}$H, respectively,
corresponding to the configuration $\lbrace \vert ++ \rangle
\langle ++ \vert, IX \rbrace$.  }   
\label{ckt1}
\end{figure}
\section{Experimental implementation of CS-QPT}
\label{nmr-expt} 
\subsection{CS-QPT of two-qubit gates}
\label{csqpt2q} 
We implemented the CS-QPT protocol for 
two, two-qubit nonlocal quantum gates,
namely, the CNOT gate and the
controlled-rotation gate.  The controlled-rotation gate is a
nonlocal gate which rotates the state of the second qubit
via $R_x(\theta)$, if the first qubit is in the state $\vert
1 \rangle $.

For two qubits the tomographically complete set of input
states is given by: $\lbrace \vert 0 \rangle, \vert 1
\rangle, \vert + \rangle, \vert - \rangle \rbrace ^{\otimes
2}$ where $\vert + \rangle = (\vert 0 \rangle + \vert 1
\rangle)/\sqrt{2} $ and $\vert - \rangle = (\vert 0 \rangle
+ i\vert 1 \rangle)/\sqrt{2} $. 
In NMR, tomographic measurements are carried out by applying
a set of unitary rotations followed by signal
acquisition\cite{long-job-2001}. The time-domain NMR signal
is recorded as a free induction decay and then Fourier
transformed to obtain the frequency spectrum, which
effectively measures the net magnetization in the transverse
($x-y$) plane.  For two NMR qubits, the tomographically
complete set of unitary rotations is given
by~\cite{li-pra-2017}: $\lbrace II, IX, IY, XX\rbrace$ where
$II$ denotes the no operation on both the qubits, $IX$
denotes no operation on the first qubit and
a $90^{\circ}$ $x$-rotation on the second qubit, $IY$ denotes
no operation on the first qubit and a $90^{\circ}$
$y$-rotation on the second qubit and $XX$ denotes a
$90^{\circ}$ $x$-rotation on both qubits.

As an illustration, the quantum circuit and corresponding
NMR implementation of the CS-QPT protocol for a two-qubit
CNOT gate is given in Fig.~\ref{ckt1}.  Fig.\ref{ckt1}(a)
depicts the general quantum circuit to acquire data for
CS-QPT and contains all possible settings corresponding to a
tomographically complete set of input quantum states and
measurements. The first block in Fig.\ref{ckt1}(a) prepares
the desired initial input state from $\vert 00 \rangle$. In
the second block the quantum process (CNOT gate in this
case) which is to be tomographed, is applied to the system
qubits and in the third block, a set of tomographic
operations are applied, followed by measurements on each
qubit.  To implement CS-QPT for any other two-qubit quantum
gate, the CNOT gate should be replaced with the desired
gate, while the remaining circuit remains unaltered.  The
first block in the Fig.\ref{ckt1}(b) represents the NMR
pulse sequence which prepares the spin ensemble in the
pseudo pure state (PPS) $\vert 00 \rangle$ and then
generates the desired input state from the $\vert 00 \rangle
$ state.  The pulse sequence corresponding to the CNOT gate
(the quantum process which is to be tomographed) is given in
the second block and finally, in the last block, the desired
set of tomographic pulses are applied and the NMR signal is
acquired.

We used $^{13}C$-enriched chloroform molecule
(Fig.\ref{ckt1}(c)) dissolved in acetone-D6 to
physically realize a two-qubit system, with the
${}^{1}$H and ${}^{13}$C spins denoting the 
first and second qubits, respectively. 
The NMR
Hamiltonian in the rotating
frame is given by: 
\begin{equation}
\mathcal{H} = - \sum_{i=1}^2 \nu_i I_{iz} +  J_{\rm CH}
I_{1z} I_{2z} \label{eq7} 
\end{equation}
where $\nu_1$, $\nu_2$ are the chemical shifts, $I_{1z}$,
$I_{2z}$ are the $z$-components of the spin angular momentum
operators of the ${}^{1}$H and ${}^{13}$C spins
respectively, and $J_{{\rm CH}}$ is the scalar coupling
constant.  We used the spatial averaging technique to
initialize the system in the PPS corresponding to $\vert 00
\rangle$, with the density matrix $\rho_{00}$ given
by~\cite{oliveira-book-07}: \begin{equation}
\rho_{00}=\frac{1}{4}(1-\eta)I_4+\eta \vert 00\rangle
\langle 00 \vert \label{eq8} \end{equation} where $\eta$
corresponds to the net spin magnetization at thermal
equilibrium, and $I_4$ is a $4 \times 4$ identity operator.
Figs.~\ref{ckt1}(d) and \ref{ckt1}(e) depict the NMR spectra
corresponding to carbon and hydrogen respectively, obtained
for the configuration $\lbrace \vert ++ \rangle, IX  \rbrace
$, where $\lbrace \vert ++ \rangle$ refers to the initial
state and $IX$ denotes the tomographic pulse set used.  The
system is prepared in the initial input state $\vert ++
\rangle$, a CNOT gate is applied, and finally the
tomographic pulse $IX$ is applied to obtain the NMR
spectrum.  For the first qubit, the area under the spectrum
is related to the density matrix elements $\rho_{24}$ and
$\rho_{13}$, while for the second qubit, the area under the
spectrum is related to the density matrix elements
$\rho_{34}$ and $\rho_{12}$.  In general, the four readout
elements of the density matrix are complex numbers; in NMR
the imaginary part of the density matrix can be calculated
by applying a $90^{\circ}$ phase shift to the spectrum
(post-processing) and then measuring the
area~\cite{long-job-2001}.  Hence a given configuration
comprises four data points (two for each qubit).  Since the
size of the full configuration space is 64 (16 states
$\times$ 4 tomographic rotations), the size of the full data
set for two qubits is $64 \times 4 = 256$. The vector
$\overrightarrow{B}^{exp}_\text{full}$ (256$\times$1
dimensional)  can be experimentally constructed by computing the area under
the spectrum for the full configuration space. One can hence
construct $\overrightarrow{B}^{exp}_m $ and the
corresponding sub-matrix $\Phi_{m}$ by randomly selecting
$m$ rows from $\overrightarrow{B}^{exp}_\text{full}$ and
$\Phi_\text{full}$ respectively, solving the optimization
problem (Eq.~(\ref{eq6})) for a reduced data set of size $m$,
and estimating the process matrix; $m$ here refers to  one
particular configuration randomly chosen from the set of all possible
256 configurations.

\subsection{CS-QPT of three-qubit gates}
\label{csqpt3q}
\begin{figure}
\includegraphics[angle=0,scale=0.95]{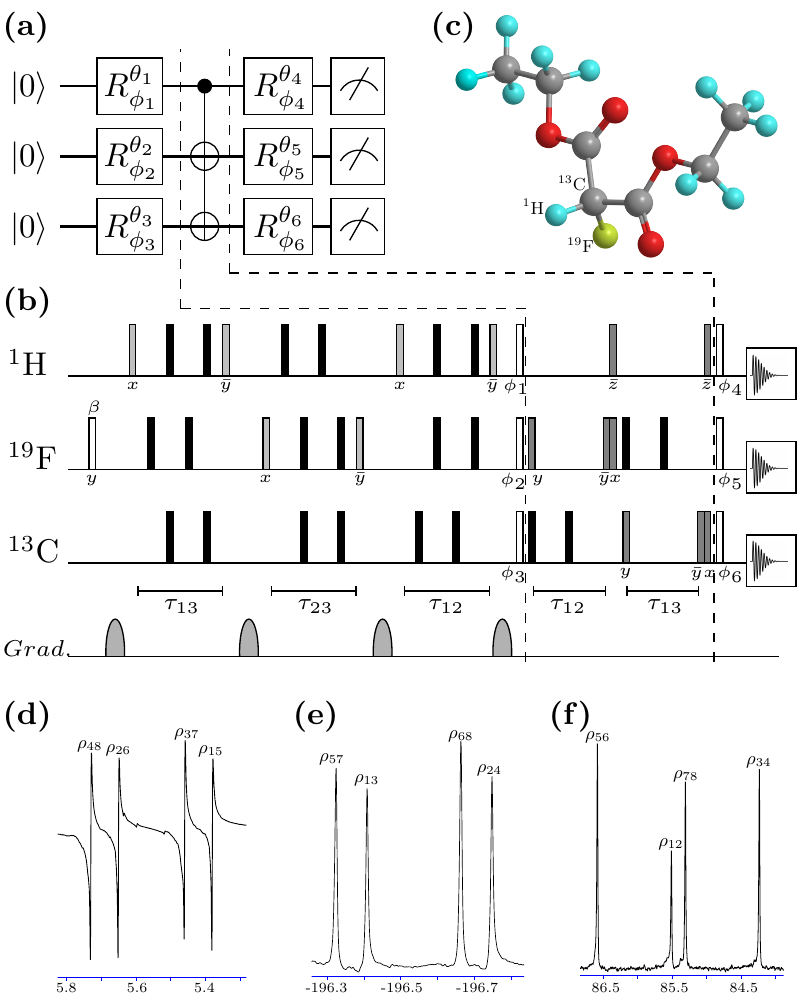}
\caption{(Color online) Quantum circuit to implement CS-QPT
of a Control-NOT-NOT ($U_{\rm {CNN}}$) gate.  The first
block prepares the desired input state while the second and
third blocks represent the quantum process corresponding to
the ($U_{\rm {CNN}}$) gate and the measurement,
respectively. (b) NMR implementation of the quantum circuit
given in panel (a). Solid black rectangles are refocusing
pulses with flip angle $180^{\circ}$, while the gray
rectangles represent pulses with flip angle $45^{\circ}$;
the corresponding rf phases are written below each pulse.
The value of $\beta$ is set to $60^{\circ}$, while the
unfilled rectangles with rf phases $\phi_1$, $\phi_2$ and
$\phi_3$ correspond to flip angles $\theta_1$, $\theta_2$
and $\theta_3$, respectively.  The black rectangles
represent pulses with flip angle $90^{\circ}$.  The unfilled
rectangles in the last block, of phases $\phi_1$, $\phi_2$
and $\phi_3$ correspond to flip angles $\theta_1$,
$\theta_2$ and $\theta_3$, respectively which implement
tomographic operations followed by measurement on each
qubit; $\tau_{ij} = \frac{ 1}{2J_{ij}}$. (c)
$^{13}C$-labeled diethyl fluoromalonate with ${}^1$H,
${}^{19}$F and ${}^{13}$C nuclei labeled as the first,
second and third qubits, respectively. NMR spectra depicted
in (d), (e) and (f) correspond to ${}^{1}$H, ${}^{19}$F and
${}^{13}$C nuclei respectively, for the configuration
$\lbrace \vert 11+ \rangle \langle 11+ \vert, XYX \rbrace$. 
}
\label{ckt3}
\end{figure}
We have implemented the CS-QPT protocol to characterize the
three-qubit controlled-NOT-NOT ($U_{{\rm CNN}}$) gate with
multiple targets, with the first qubit being denoted the control
qubit,
while the other two qubits are the target qubits.  The 
controlled-NOT-NOT gate
can be decomposed using two CNOT gates as:~$U_{{\rm CNN}} \equiv$
CNOT$_{13}.$CNOT$_{12}$, and is widely used in encoding
initial input states in error correction codes, fault
tolerant operations~\cite{egan-arxiv-2021,shor-pra-1995} and
in the preparation of three-qubit maximally entangled
states~\cite{mooney-arxiv-2021,singh-pra-2018,dogra-pra-2015}.  

The NMR Hamiltonian for three qubits
in the rotating frame is given by:
\begin{equation}
\mathcal{H} = - \sum_{i=1}^3 \nu_i I_{iz} + \sum_{i,j=1
(i \ne j)}^3 J_{ij} I_{iz} I_{jz} \label{ham3q} 
\end{equation}
where the indices $i,j$ label the qubit and
$\nu_i$ denotes the respective chemical shift. The 
quantity $J_{ij}$ denotes the scalar coupling strengths
between the $i$th and $j$th qubits, while $I_{iz}$ represents
the $z$-component of the spin angular momentum of the $i$th qubit. We have
used ${}^{13}$C-labeled diethyl fluoromalonate
(Fig.\ref{ckt3}(c)) dissolved in acetone-D6 to physically
realize a three-qubit system, with the
${}^{1}$H, ${}^{19}$F and ${}^{13}$C nuclei being labeled as the first,
second and third qubits, respectively. State initialization is
performed by preparing the system in the PPS
$\vert 000 \rangle$ via the spatial averaging
technique with the corresponding density matrix being given
by: 
\begin{equation} 
\rho_{000} = (\frac{1-\epsilon}{8})I_8
+ \epsilon \vert 000 \rangle \langle 000 \vert \label{pps3}
\end{equation} 
where $\epsilon \approx 10^{-5}$ represents
the net thermal magnetization and $I_8$ is the 8$\times$8 identity
operator.

For a three-qubit system, the tomographically complete set
of input states is given by: $\lbrace \vert 0 \rangle, \vert
1 \rangle, \vert + \rangle, \vert - \rangle \rbrace
^{\otimes 3}$ where $\vert + \rangle = (\vert 0 \rangle +
\vert 1 \rangle)/\sqrt{2} $ and $\vert - \rangle = (\vert 0
\rangle + i\vert 1 \rangle)/\sqrt{2} $ and the tomographically
complete set of unitary rotations is given
by:~$\lbrace III, IIY, IYY, YII, XYX, XXY,
XXX \rbrace$~\cite{li-pra-2017}. 
The quantum circuit and the corresponding NMR
pulse sequence to perform CS-QPT for the three-qubit gate $U_{\rm
CNN}$ is given in Fig.~\ref{ckt3}. The first block in
Fig.\ref{ckt3}(a) represents the input state
preparation while the second block represents the
application of quantum gate $U_{\rm{CNN}}$ (i.e. quantum
process which is to be tomographed),
and tomographic unitary rotations are
applied in the last block, followed by measurement on each qubit. 
Fig.\ref{ckt3}(b) represents the corresponding NMR
implementation of quantum circuit given in the
Fig.\ref{ckt3}(a).  The spatial averaging
techniques are used in the first block~\cite{singh-pra-2019} 
to initialize system in the desired PPS,
followed by the application of spin-selective rf
pulses to prepare the desired input state. 
In the second
block the pulse sequence corresponding to $U_{\rm{CNN}}$ is
applied on the input state and in the last block after
application of tomographic pulses, the signal 
of the desired nucleus is recorded.
The NMR spectra corresponding to ${}^1$H, ${}^{19}$F and
${}^{13}$C are given in
Figs.~\ref{ckt3}(d), (e) and (f), respectively,
for the configuration $\lbrace \vert
11+ \rangle \langle 11+ \vert , XYX \rbrace $, \ie the
input state  $ \vert 11+ \rangle \langle 11+ \vert $ is
prepared, evolved under the quantum process corresponding
to $U_{\rm{CNN}}$, the tomographic set of pulses $XYX$ is
applied, and finally the NMR signal is recorded. 
For the first qubit (${}^{1}$H)
the area under the four spectral lines 
correspond to the
density matrix elements
$\rho_{48}$, $\rho_{26}$, $\rho_{37}$ and $\rho_{15}$,
for the second qubit (${}^{19}$F)
the area under the four spectral lines 
correspond to the
density matrix elements
$\rho_{57}$, $\rho_{13}$, $\rho_{68}$ and $\rho_{24}$, while
for the third qubit (${}^{13}$C),
the area under the four spectral lines 
correspond to the
density matrix elements
$\rho_{56}$, $\rho_{12}$, $\rho_{78}$ and 
$\rho_{34}$, respectively. 
For a three-qubit
system there are 12 experimental data points (4 per
qubit) for a given configuration and the total number
configurations are 448 (64 input states $\times$ 7
tomographic unitary operations) which yields the
$\overrightarrow{B}^{exp}_\text{full}$ of size = 5376 (448
configurations $\times $ 12 data points per configuration).
One can construct $\overrightarrow{B}^{exp}_\text{m}$ by
randomly selecting $m$ number of rows from
$\overrightarrow{B}^{exp}_\text{full}$, and using the
corresponding coefficient matrix $\Phi_{m}$ one can solve
the optimization problem (Eq.~(\ref{eq6})), and construct 
the process matrix for a reduced data
set of size $m$.

\subsection{CS-QPT of two-qubit processes in 
a three-qubit system}
\label{sec3.3}
In order to experimentally implement a two-qubit CNOT gate
in a multi-qubit system, one needs to allow the two system
qubits to interact with each other \ie, let them evolve under
the internal coupling Hamiltonian for a finite time. In
reality, this is non-trivial to achieve experimentally, as
during the evolution time the other qubits are also
continuously interacting with system qubits, and one has to
``decouple'' the system qubits from the other qubits. In the
language of NMR, this is  referred to as refocusing of the
scalar $J$-coupling.  

To implement a two-qubit CNOT gate we need four single-qubit
rotation gates and one free evolution under the internal
coupling Hamiltonian (Fig.~\ref{ckt1}).  The single-qubit
rotation gates are achieved by applying very short duration
rf pulses of length $\approx 10^{-6}$ s, while the time
required for free evolution under the coupling Hamiltonian
is $\approx 10^{-3}$ s. The quality of the experimentally
implemented quantum gate depends on the time required for
gate implementation, which for the two-qubit CNOT gate, is
primarily determined by the free evolution under the
coupling Hamiltonian.
We use the CS-QPT protocol to
efficiently characterize three coupling evolutions
corresponding to $U^J_{ij}$ of the form: 
\begin{equation}
U^J_{ij}(t) = e^{-i 2 \pi J_{ij} I_{iz} I_{jz} t} \label{hf}
\end{equation} 
where the indices $i$ and $j$ label the qubits
and $J_{ij}$ is the 
strength of the scalar
coupling between the $i$th and the $j$th qubit; 
for the CNOT gate,
$t = \vert \frac{1}{2 J_{ij}} \vert $. 
A three-qubit system is continuously evolving
under all the three $J_{ij}$ couplings, so in order to let a
subsystem of two qubits effectively evolve under one 
of these
couplings, we have to refocus all the other $J$-couplings.
For example, consider the two-qubit subsystem of the $i$th and  $j$th
qubit with the effective evolution $U_{ij}^{J}(t)$ given by:
\begin{equation}
U_{ij}^{J}(t) = U_{\rm{int}}(\frac{t}{2}) R_{x}^k(\pi)
U_{\rm{int}}(\frac{t}{2}) R_{x}^k(-\pi)
\end{equation}
where $R_{x}^k(\pm \pi)$ is an $x$-rotation on the $k$th
qubit by an angle $\pm \pi$ and $U_{{\rm int}}(\frac{t}{2})$
is the unitary operator corresponding to free evolution for
a duration $\frac{t}{2}$ under the internal Hamiltonian
$\mathcal{H}_{{\rm int}} = \sum_{i,j=1, i>j}^3 J_{ij} I_{iz}
I_{jz}$.  The procedure for tomographic reconstruction of
the reduced two-qubit density matrix from the full
three-qubit density matrix is given in Table.~\ref{table1}.
We were able to successfully characterize all three
$U_{ij}^{J}(t)$ via the CS-QPT method and constructed the
corresponding process matrices, using a heavily reduced data
set of size $\approx 20$, with experimental fidelities $>
0.94$.  Using the information given in Table~\ref{table1},
one can efficiently characterize a general quantum state as
well as the dynamics of a two-qubit subsystem in a
three-qubit system, wherein the experimental data is
acquired by measuring only the two qubits under
consideration; hence the complete set of input states and
tomographic rotations required are the same as for the
two-qubit protocol described in Section~\ref{csqpt2q}.

\begin{table}[h!] \caption{\label{complexity3} Relation
between the readout positions 
of the reduced density matrix of the subsystem ($\rho_{ij}^{\prime}$)
and the readout positions $\rho_{mn}$.}
\begin{ruledtabular}
\begin{tabular}{c | c c c c}
~Subsystem ~& \multicolumn{4}{c}{Readout positions of the reduced density
matrix}~ \\
~~ & ~$\rho_{24}'$ & ~$\rho_{13}'$ & ~$\rho_{34}'$ &
~$\rho_{12}'$ ~~~~\\ \colrule
~${}^{1}$H+${}^{19}$F~ & ~$\rho_{48}+\rho_{37}$ &
~$\rho_{26}+\rho_{15}$ & ~$\rho_{57}+\rho_{68}$ &~
$\rho_{13}+\rho_{24}$ ~~~ \\
~${}^{1}$H+${}^{13}$C~ & ~$\rho_{48}+\rho_{26}$ &
~$\rho_{37}+\rho_{15}$ & ~$\rho_{56}+\rho_{78}$ &~
$\rho_{12}+\rho_{34}$ ~~~ \\
~${}^{19}$F+${}^{13}$C~ & ~$\rho_{68}+\rho_{24}$ &~
$\rho_{57}+\rho_{13}$ & ~$\rho_{78}+\rho_{34}$ &~
$\rho_{56}+\rho_{12}$ ~~
\label{table1}
\end{tabular}
\end{ruledtabular}
\end{table}
\begin{figure*}[t]
\includegraphics[angle=0,scale=1]{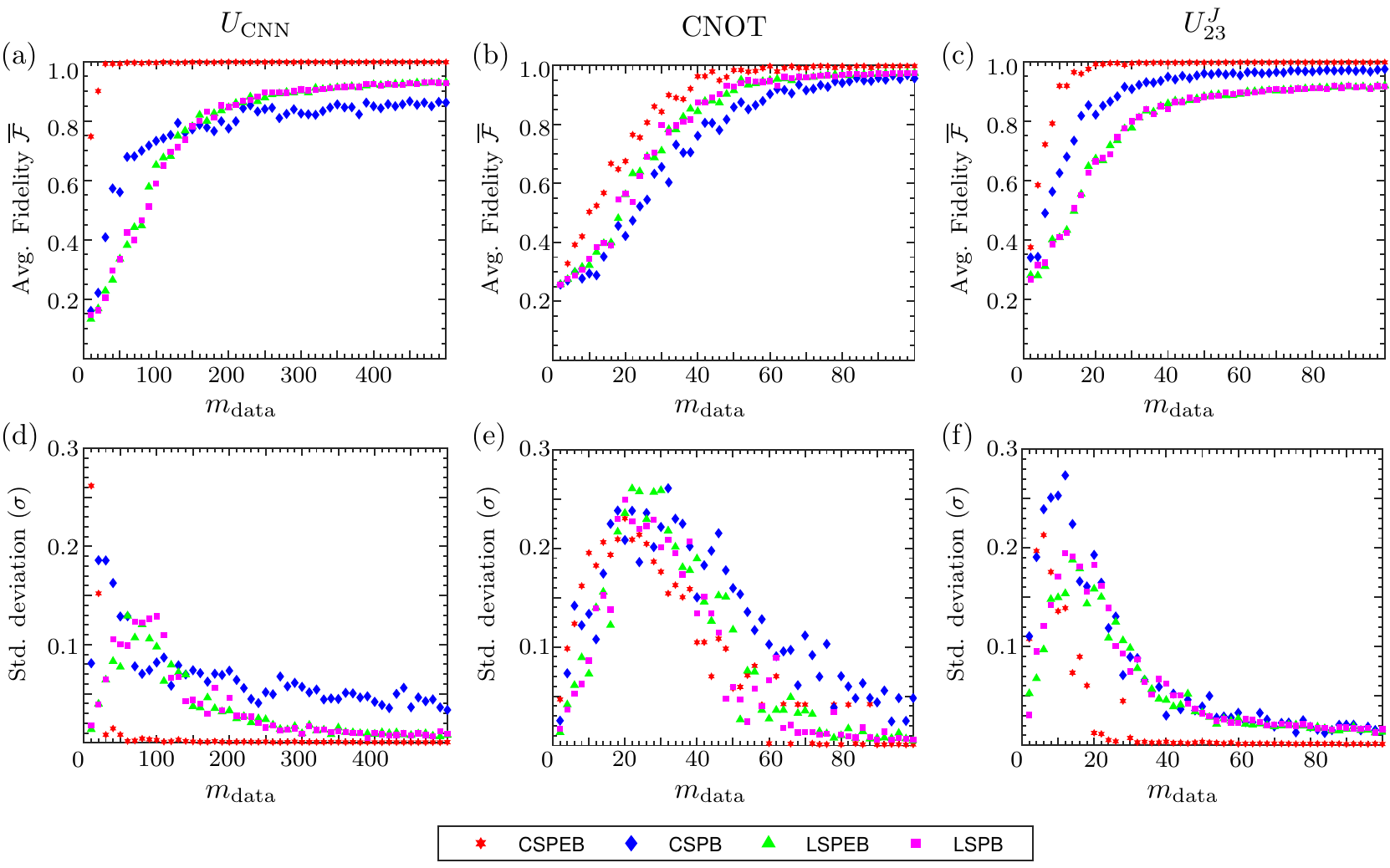} 
\caption{(Color online)
The top panel represents the average gate fidelity
$\overline{\mathcal{F}}$ corresponding to (a) a three-qubit
$U_{\rm {CNN}}$ gate, (b) a CNOT gate and (c) $U^J_{23}$
against the number of data points $m_{\text{data}}$ on the
$x$-axis.  The bottom panel represents the standard
deviation in average fidelity $\sigma$, corresponding to (d)
a three-qubit $U_{\rm {CNN}}$ gate, (e) a CNOT gate, and (f)
$U^J_{23}$, plotted against the number of data points
$m_{\text{data}}$ on the $x$-axis.  The data points in red,
blue, green and pink correspond to CS-PEB, CS-PB, LS-PEB and
LS-PB methods, respectively. The CS-PEB method shows the
best performance for all three quantum gates.
}
\label{plots}
\end{figure*}
\begin{table*}[t]
\caption{\label{complexity3}
The minimum value of $m_{\rm{data}}$ at which the experimental average gate
fidelity $\overline{\mathcal{F}}$ turns to be $> 0.9$ is computed (alongwith
the standard deviation $\sigma$) for different quantum processes, via the
CS-PEB, CS-PB, LS-PEB and LS-PB methods.
}
\begin{ruledtabular}
\begin{tabular}{r|c c c|c c c|c c c|c c c }
 & \multicolumn{3}{c|}{CS-PEB} & \multicolumn{3}{c|}{CS-PB}
& \multicolumn{3}{c|}{LS-PEB} & \multicolumn{3}{c}{LS-PB} \\
Gate & $m_{\rm {data}}$ &$\overline{\mathcal{F}}$ & $\sigma$
& $m_{\rm {data}}$ &$\overline{\mathcal{F}}$ & $\sigma$&
$m_{\rm {data}}$ &$\overline{\mathcal{F}}$ & $\sigma$&
$m_{\rm {data}}$ &$\overline{\mathcal{F}}$ & $\sigma$  ~~~\\
\colrule
$U_{\rm{CNN}}$ & 30 & 0.9920 & ~0.0081~ & - & - & - & 320 &
0.9109 & ~0.0123~ & 290 & 0.9006 & 0.0147 ~~~\\
CNOT & 44  & 0.9798  &  0.0701 & 62  &  0.9203  & ~0.0905~ &
52    & 0.9514   &    0.0263 & 48  &  0.9308  &
0.0475~~~\\
C-$R_x^{\pi}$ & 48  &  0.9728   &   0.0797 & 58  &   0.9068
& 0.0746 & 48  &  0.9332  &   0.0805 & 52  &  0.9503  &
0.0504 ~~~\\
$U^J_{12}$ & 14  &  0.9549  &    0.0963 & 24  &  0.9464
& 0.0468 & 32 &  0.9075  &  0.0459 & 34  &  0.9071  &
0.0561 ~~~\\
$U^J_{23}$ & 14  &  0.9641  &   0.0734 & 28  &  0.9145  &
0.0710 & 66  &  0.9019  &  0.0217 & 68  &  0.9048  &
0.0198~~~\\
$U^J_{13}$ & 18  &  0.9417 &  0.0980 & 38  &  0.9067 & 0.0695
 & -  & -  & -  & -  &  - & -  
\label{table2}
\end{tabular}
\end{ruledtabular}
\end{table*}
\subsection{Comparison of CS-QPT and LS-QPT protocols}
\label{sec3.4}
\begin{table}[h!] 
\centering
\caption{\label{fid} Experimental quantum process
fidelities obtained via CS and LS methods using 
the full data set $m^{\rm{full}}_{\rm{data}}$.}
\begin{tabular}{r c c c c}
\hline
Gate & ~~CS-PEB~~& ~~CS-PB~~ & ~~LS-PEB~~& ~~LS-PB\\
\hline $U_{\rm{CNN}}$ & 0.9980 & 0.8877 & 0.9542 &~ 0.9542\\ 
CNOT & 0.9984 & 0.9843 & 0.9817 &~ 0.9817\\ 
C-$R^{\pi}_x$ & 0.9980 & 0.9744 & 0.9831 &~ 0.9831\\
$U^J_{12}$ & 0.9967 & 0.9894 & 0.9819 &~ 0.9819 \\
$U^J_{23}$ & 0.9976 &  0.9793 & 0.9273 &~ 0.9273 \\
$U^J_{13}$ & 0.9895 & 0.9710 & 0.8942 &~ 0.8942 \\
\hline 
\end{tabular}
\end{table}
The fidelity of the experimentally estimated $\chi_{{\rm exp}}$
is computed using the
measure\cite{zhang-prl-2014}:
\begin{equation}
{\mathcal F}(\chi^{}_{\rm exp},\chi^{}_{\rm ideal})=
\frac{|{\rm Tr}[\chi^{}_{\rm exp}\chi_{\rm ideal}^\dagger]|}
{\sqrt{{\rm Tr}[\chi_{\rm exp}^\dagger\chi^{}_{\rm exp}]
{\rm Tr}[\chi_{\rm ideal}^\dagger\chi^{}_{\rm ideal}]}}
\label{eq9}
\end{equation} 
where
$\chi_{\text{ideal}}$ is the theoretically constructed process
matrix,
and as $\chi_{\rm{exp}} \rightarrow
\chi_{\rm{ideal}}$, ${\mathcal F}(\chi^{}_{\rm
exp},\chi^{}_{\rm ideal}) \rightarrow 1$.

We performed QPT of several two- and three-qubit quantum gates using both
CS-QPT and LS-QPT protocols on a reduced data set.  The CS-QPT method was
implemented for the PEB and PB basis sets.  For a two-qubit system
$m_{\text{data}}^{\rm full} = 256$,  while for a three-qubit system,
$m_{\text{data}}^{\rm full} = 5376$,  where $m_{\text{data}}^{\rm full}$
denotes the size of the full data set obtained using the complete set of input
states and tomographic rotation operators for two and three qubits as given in
Sections~\ref{csqpt2q} and \ref{csqpt3q}, respectively.  In the PEB basis,
$\chi_{\text{ideal}}$ is maximally sparse for all unitary quantum gates, while
in the PB basis, $\chi_{\text{ideal}}$ corresponding to the two-qubit CNOT,
controlled-$R_{x}^{\pi}$ and $U^J_{ij}$ gates have 16, 16 and 4 non-zero
elements, respectively (out of a total of 256 elements).  For the three-qubit
gate $U_{\rm CNN}$, $\chi_{\text{ideal}}$ has 16 non-zero elements (out of a
total of 4096 elements).

The performance of the CS-QPT method was compared  with the
LS-QPT method for six different quantum processes corresponding to: (i) a
three-qubit $U_{\rm CNN}$ gate, (ii) a two-qubit CNOT gate, (iii) a
controlled-$R_x^{\pi}$ rotation (iv) $U^J_{23}$, (v) $U^J_{13}$ and (vi)
$U^J_{12}$, of which the results of the quantum process corresponding to (a) a
three-qubit $U_{\rm CNN}$ gate, (b) a two-qubit CNOT gate and (c) $U^J_{23}$,
are displayed in Fig.~\ref{plots}. The top panel in Fig.~\ref{plots} represents
the average gate fidelity $\overline{\mathcal{F}}$ 
plotted against $m_{\text{data}}$, while the
bottom panel represents the standard deviation $\sigma$ in average gate fidelity
plotted against $m_{\text{data}}$. The average gate fidelity is
obtained using the average process matrix 
estimated via the LS and CS algorithm in the PEB
and PB bases. 
The plots in red and blue color represent the results of 
the CS-QPT method implemented
in PEB and PB basis respectively, while the 
plots in green and pink color represent
the results of the LS-QPT 
method implemented in the PEB and PB basis, respectively. The average fidelity
and the value of $\sigma$ is computed by implementing the CS-QPT and LS-QPT
protocols 50 times for randomly selected $m_{\text{data}}$ number of data
points, and $\sigma$ is calculated from: 
\begin{equation}
\sigma = \sqrt{\frac{\sum_{i=1}^{N}
(\mathcal{F}_i-\overline{\mathcal{F}})^2}{N-1}} \label{eq10}
\end{equation} 
where $N=50$ and $\overline{\mathcal{F}}$ is
the average fidelity.

The plots in the first column of Fig.~\ref{plots} correspond to the
three-qubit gate $U_{\rm{CNN}}$, where Fig.~\ref{plots}(a) depicts the
accuracy, while Fig.~\ref{plots}(d) gives the precision in characterizing
$U_{\rm{CNN}}$, for a given value of $m_{\text{data}}$. Similarly, the second
and third columns in Fig.~\ref{plots} represent the experimental results
corresponding to the CNOT gate and the $U^J_{23}$ quantum 
process, respectively. The plots
corresponding to the two-qubit controlled-rotation gate (C-R$_{x}^{\pi}$) is
similar to the CNOT gate, while the plots corresponding to $U^J_{13}$ and
$U^J_{12}$ are similar to  $U^J_{23}$ (plots not shown). As seen from
Fig.~\ref{plots}, the CS-QPT method implemented in the PEB basis, performs
significantly better than the LS-QPT and the CS-QPT methods implemented in the
PB basis, for all the quantum processes considered.  The performance of the
LS-QPT method is independent of the choice of basis operators.  On the other
hand, the CS-QPT method may yield a lower fidelity as compared to the LS-QPT
method, if the basis operators are not properly chosen.  Using a reduced data
set, the overall performance for the three-qubit gate $U_{\rm{CNN}}$ is CS-PEB
$>$ CS-PB $>$ LS-PEB $\approx$ LS-PB, while for the two-qubit CNOT and
C-R$_x^{\pi}$ gates, CS-PEB $>$ LS-PEB$\approx$ LS-PB $>$ CS-PB.  For the
two-qubit $U_{ij}^J$ processes, CS-PEB $>$ CS-PB $>$ LS-PEB $\approx$ LS-PB.

For the two-qubit CNOT and C-R$_x^{\pi}$ gates, the LS algorithm performs
better than the CS algorithm in the PB basis for all values of $m_{\rm{data}}$,
while for the three-qubit $U_{\rm{CNN}}$ gate,  the LS algorithm performs
better than the CS algorithm in the PB basis for $m_{\rm{data}} \geq 160$,
which clearly shows the importance of selecting an appropriate operator basis
set while implementing the CS algorithm. 
The plots given in Fig.~\ref{plots} provide information about
the experimental complexity of the CS and LS algorithms \ie the
number of experiments required in each case to characterize a
given quantum process.
We note here in passing that the
standard deviation in average fidelity ($\sigma$) is not monotonic.  For small
values of $m_{\rm {data}}$, the process of randomly selecting $m_{\rm {data}}$
data points to estimate the process matrix is more likely to lead to a lower
fidelity and higher values of the standard deviation $\sigma$, and hence lower
precision.  For the two-qubit CNOT and controlled-R$_x^{\pi}$ gates, $\sigma$
has a maximum  around $m_{\rm {data}}\approx 20 $, while for the $U_{CNN}$,
$U^J_{23}$, $U^J_{13}$ and $U^J_{12}$ quantum processes, $\sigma$ is maximum
around $m_{\rm {data}}\approx 10 $. For all the cases, the CS-PEB method yields
better precision as compared to the CS-PB, LS-PEB and LS-PB methods.

The experimentally obtained minimum value of  $m_{\rm {data}}$  at which the
experimentally computed average gate fidelity is $> 0.9$ is given in
Table~\ref{table2}, for all the quantum processes.  For the three-qubit
$U_{\rm{CNN}}$ gate, we experimentally obtained
$\overline{\mathcal{F}}_{\rm{CS-PEB}} = 0.9920 \pm 0.0081$ for a reduced data
set of size $m_{\rm{data}} = 30$.  For the two-qubit CNOT and
control-$R_x^{\pi}$ gates, $\overline{\mathcal{F}}_{\rm{CS-PEB}} \geq  0.9790
\pm 0.0701 $ and $\overline{\mathcal{F}}_{\rm{CS-PEB}} \geq 0.9729 \pm 0.0797 $
for $m_{\rm{data}} \geq 44$ and $m_{\rm{data}} \geq 48$, respectively. The
reduced data set is $\approx 5$ times smaller than the full data set, which
implies that the experimental complexity is reduced by $\approx 80 \%$ as
compared to the standard QPT method. Furthermore, for all the two-qubit quantum
processes corresponding to $U_{ij}^J$, $\overline{\mathcal{F}}_{\rm{CS-PEB}}
\geq  0.9417 \pm 0.0980 $ for $m_{\rm{data}} \geq 18$. This reduced data set is
$\approx 12$ times smaller than the full data set which implies that the
experimental complexity in these cases is reduced by $\approx 92 \%$ as
compared to the standard QPT method.  
\section{Concluding Remarks}
\label{concl} 
We designed a general quantum circuit to acquire experimental data compatible
with the CS-QPT algorithm.  The proposed quantum circuit can also be used for
other experimental platforms and can be extended to higher-dimensional systems.
We successfully demonstrated the efficacy of the CS-QPT protocol for various
quantum processes corresponding to the three-qubit $U_{\rm{CNN}}$ gate,
two-qubit CNOT and controlled-rotation gates and several two-qubit $U_{ij}^J$
unitary operations.  Our experimental comparison of the CS-QPT and LS-QPT
schemes demonstrate that the CS-QPT protocol is far more efficient, provided
that the process matrix is maximally sparse and that an appropriate operator
basis is chosen.  

Standard QPT protocols do not have access to prior information about the
intended target unitary and hence require a large number of parameters to
completely characterize the unknown quantum process.  CS methods can be used to
dramatically reduce the resources required to reliably estimate the full
quantum process, in cases where there is substantial prior information
available about the quantum process to be characterized.  Since the CS-QPT
method is uses fewer resource and is experimentally viable, it can be used to
characterize higher-dimensional quantum gates and to validate the performance
of large-scale quantum devices.

\begin{acknowledgments}
All the experiments were performed on a Bruker Avance-III
600 MHz FT-NMR spectrometer at the NMR Research Facility of
IISER Mohali.  Arvind acknowledges financial support from
DST/ICPS/QuST/Theme-1/2019/Q-68.
K.~D. acknowledges financial support from
DST/ICPS/QuST/Theme-2/2019/Q-74.
\end{acknowledgments}

%

\end{document}